# Unidirectional guided-wave-driven metasurfaces for arbitrary wavefront control

Shiqing Li[1], Kosmas L. Tsakmakidis[2]*, Tao Jiang[3], Qian Shen[1], Hang Zhang[1], Jinhua Yan[1], Shulin Sun[4] and Linfang Shen[1]*

[1]*Department of Applied Physics, Zhejiang University of Technology, Hangzhou 310023, China.*
[2]*Section of Condensed Matter Physics, Department of Physics, National and Kapodistrian University of Athens Panepistimioupolis, Athens GR-157 84, Greece.*
[3] *Yangtze Delta Region Institute (Huzhou), University of Electronic Science and Technology of China, Huzhou 313001, China.*
[4]*Shanghai Engineering Research Centre of Ultra Precision Optical Manufacturing, Department of Optical Science and Engineering, School of Information Science and Technology, Fudan University, Shanghai 200433, China.*

# Abstract

Metasurfaces, composed of subwavelength electromagnetic microstructures, known as meta-atoms, are capable of reshaping the wavefronts of incident beams in desired manners, making them great candidates for revolutionizing conventional optics. However, the requirement for external light excitation and the resonant nature of meta-atoms make it difficult to fully integrate metasurfaces on-chip or to control wavefronts *at deep-subwavelength scales*. Here, we introduce the concept and design of a new class of metasurfaces, driven by *unidirectional* guided waves, and being capable of arbitrary wavefront control based on the unique dispersion properties of unidirectional guided waves rather than resonant meta-atoms. Upon experimentally demonstrating the feasibility and practicality of the unidirectional nature of our designs in the microwave regime, we numerically validate this new principle through the design of several microwave meta-devices using metal-air-gyromagnetic unidirectional surface magnetoplasmons, agilely converting unidirectional guided modes into the wavefronts of 3D Bessel beams, focused waves, and controllable vortex beams. We also numerically demonstrate *sub-diffraction* focusing, which is currently beyond the capability of conventional metasurfaces. Furthermore, we directly show how these concepts can be transferred to the terahertz regime, and discuss their feasibility in the optical domain, too. Based on this *nonresonant* (that is, broadband) mechanism and on standard plasmonic platforms, our metasurfaces can be integrated on-chip, enabling the manipulation of electromagnetic waves on deep subwavelength scales and over wide frequency ranges, thereby opening up new opportunities for applications in communications, remote sensing, displays, and so forth.

*Corresponding authors: Linfang Shen, lfshen@zjut.edu.cn; Kosmas L. Tsakmakidis, ktsakmakidis@phys.uoa.gr.

**Introduction**

Light can be confined at the nanoscale through coupling with surface plasmons[1]. The 'two-dimensional' nature of the resulting surface plasmon polaritons (SPPs) offers significant flexibility in engineering photonic integrated circuits for optical communications and optical computing[2-4]. However, the backscattering of SPPs by disorders, defects and structural imperfections, due to the reciprocity of plasmonic platforms, limits their applications in optical systems. Developing nonreciprocal plasmonic platforms[5] that enable unidirectional SPP propagation is, therefore, of great importance. Such unidirectional SPPs occur in nonreciprocal plasmonic platforms made of magnetized semiconductors in the terahertz regime[6] or magnetized metals in the visible regime[7], possessing gyroelectric anisotropy induced by an external magnetic field. These unidirectional SPPs, known as unidirectional surface magnetoplasmons (USMPs) for several decades, have recently regained interest in the context of topological electromagnetics[7-15]. It has been revealed that true USMPs are topologically protected[12,13] and thus robust against nonlocal effects. These USMPs can be immune to backscattering at disorders due to the absence of a back-propagating mode in the system. In the microwave domain, nonreciprocal materials such as yttrium-iron-garnet (YIG) generally exhibit gyromagnetic anisotropy under an external magnetic field, which can also support unidirectional surface polaritons possessing almost the same guiding properties as USMPs[16-18]. Thus, these low-frequency unidirectional electromagnetic (EM) modes are also referred to as USMPs. Robust USMPs provide a fundamental mechanism for realizing new classes of optical devices that are impossible using conventional reciprocal EM modes, such as optical cavities that overcome the time-bandwidth limit[19].

Relying on the bandgap of nonreciprocal materials themselves, USMPs easily attain broad bandwidth and simultaneously exhibit unique dispersion properties. For example, in the microwave range, the dispersion curve of USMPs in YIG materials can monotonically grow across the entire light cone in air. This implies that USMPs can smoothly convert from waves with positive velocities to waves with negative velocities without any frequency gap while they maintain the sign of their group velocities. As the dispersion of USMPs is closely related to the structural details and, hence, can be flexibly tailored through controlling the structural parameters. For USMP at a fixed frequency, the phase constant can be tuned over the range [−$k_0$, $k_0$] (where $k_0$ is the free-space wavenumber). Based on this unparalleled phase controllability, it is interesting to investigate whether USMPs may offer a completely new mechanism for designing metasurfaces, one that would have no resonant nature. Metasurfaces[20-22] are capable of reshaping the wavefronts of incident beams in desired manners, and they are thought to be great candidates for revolutionizing conventional optics. Relying on engineered optical resonators, known as meta-atoms, metasurfaces can locally provide abrupt phase shifts at subwavelength intervals to tailor the phase of incident waves, making them exhibit predetermined functions for transmitted or reflected waves based on the Huygens principle. These metasurfaces can provide unparalleled control over electromagnetic (EM) waves to realize complex free-space functions, such as beam deflection[23,24], focusing[25-27], generation of orbital angular momentum beams[28], and holograms[29]. However, most metasurfaces are driven by free-space waves, making their on-chip integration challenging. Moreover, based on the resonant mechanism of the meta-atoms, the designed wavelengths of metasurfaces are closely related to the sizes of the meta-atoms, making it difficult to shape wavefronts on deep subwavelength scales. These metasurfaces also generally operate within a narrow frequency range.

In this paper, we develop a new class of metasurfaces that are driven by USMPs (Fig. 1) based on their unparalleled phase controllability. The subwavelength-sized unit cells of USMP-driven metasurfaces support two types of USMPs, and the conversion between them creates a directional radiation that extracts guided waves into free space and molds them into desired light fields. Based on the robustness of USMPs, the proposed metasurfaces would be utterly immune to backscattering from surface-roughness. Besides, multiple metasurfaces can be directly connected to achieve different free-space functions simultaneously. In contrast to existing guide wave-driven metasurfaces[30-32], our USMP-driven metasurfaces provide desired spatial phase profiles for extracted waves directly from the propagation of USMPs, thus do not need any meta-atoms to induce abrupt phase shifts. Although plasmonic metasurfaces based on Bragg interferences also work without meta-atoms[33-38], the wavefront can only be controlled at wavelength scale due to the limited phase controllability of conventional SPPs. In contrast, USMP-driven metasurfaces can even manipulate EM waves on deep subwavelength scale. The developed technology paves exciting ways for building multifunctional USMP-driven meta-devices with flexible access to free space, offering advantages such as ease of fabrication, reconfiguration, compatible with on-chip technology. The technology can excite many related applications such as communications, remote sensing, and virtual reality display.

## Results

### Phase controllability of USMPs

We first investigate nonreciprocal waveguide systems that can support USMPs. The material configuration plays a critical role in determining the properties of USMPs, and we consider two types of nonreciprocal waveguides, as illustrated in Fig. 2a: a YIG-dielectric-metal layered structure (type-I waveguide) and a YIG-air layered structure (type-II waveguide). Both types of waveguides can support USMPs at microwave frequencies. For the type-I waveguide, we assume that the dielectric has a relative permittivity close to 1 (e.g., foam). In both waveguides, the YIG ($\varepsilon_m = 15$)[39] possesses remanent magnetization $\hat{M}_m = \hat{y} M_m$ that breaks the time-reversal symmetry of the system, enabling the existence of USMP in the absence of external magnetic field. The remanence induces gyromagnetic anisotropy in the YIG, with the permeability tensor taking the form

$$\mu_m = \begin{bmatrix} \mu_1 & 0 & -i\mu_2 \\ 0 & 1 & 0 \\ i\mu_2 & 0 & \mu_1 \end{bmatrix}, \tag{1}$$

where $\mu_1 = 1$, and $\mu_2 = \omega_m/\omega$ ($\omega$ is the angular frequency). Here, $\omega_m = 2\pi f_m = \mu_0 \gamma M_m$ is the characteristic circular frequency ($\mu_0$ is the vacuum permeability, $\gamma$ is the gyromagnetic ratio). Surface magnetoplasmons (SMPs) in the type-I waveguide have a dispersion property closely depending on the dielectric-layer thickness ($d$), which is described by the dispersion relation

$$\alpha_r \mu_v + \left( \alpha_m + \frac{\mu_2}{\mu_1} k \right) \tanh(\alpha_r d) = 0, \tag{2}$$

where $k$ is the phase constant, $\alpha_m = \sqrt{k^2 - \varepsilon_m \mu_v k_0^2}$ with $\mu_v = \mu_1 - \mu_2^2/\mu_1$, and $\alpha_r = \sqrt{k^2 - k_0^2}$ for $|k| > k_0$ and $\alpha_r = -i\sqrt{k_0^2 - k^2}$ for $|k| \leq k_0$. The dispersion relation for the type-II waveguide can be directly obtained from Eq. (2) by letting $d = \infty$, which gives

$$\alpha_r \mu_v + \alpha_m + \frac{\mu_2}{\mu_1} k = 0. \tag{3}$$

The linear terms with respect to $k$ in Eqs. (2), (3) imply that the two waveguides are intrinsically non-reciprocal, allowing for the possibility of unidirectional wave proapgation.

The dispersion equation (2) for the type-I waveguide was numerically solved, yielding results displayed in Fig. 2b, where various values of $d$ were analyzed. The dispersion relation for SMPs in the type-I waveguide exhibits a single branch that extends across the entire light cone in air. The positive slop of the dispersion curve unambiguously establishes the USMP (type-I USMP) nature of SMPs. Moreover, the dispersion curve descends as $d$ increases, allowing for tailored control of the phase constant $k$ by adjusting the thickness of the dielectric layer. In all cases, type-I USMPs possess an asymptotic frequency of $\omega_{smp} = 0.5\omega_m$, which sets a lower-frequency cutoff. However, robust unidirectional propagation (RUP) is achieved exclusively in the ferrite bandgap, where no backward-propagating mode exists. Consequently, robust type-I USMP occurs within the frequency range $[0.5\omega_m, \omega_m]$. The dispersion relation (3) for the type-II waveguide is also depicted in Fig. 2b, showing two asymmetric branches: one with $k < 0$ and the other with $k > 0$. Both dispersion branches lie beyond the light cone in air. The branch with $k < 0$ has an asymptotic frequency of $\omega_{smp} = 0.5\omega_m$, setting its upper-frequency cutoff. As a result, USMP (type-II USMP) occurs in the range of $\omega_{smp} > 0.5\omega_m$. The type-II USMP lacks robustness when encountering obstacles or disorder, resulting in partial scattering into free space. Clearly, within the range of our concern $[0.5\omega_m, \omega_m]$, both types of waveguides support USMPs. In contrast to type-II USMP with $k > k_0$, type-I USMP possesses a tunable phase constant $k$ range over $[-k_0, k_0]$ by varying $d$. For instance, at $f = 0.75 f_m$, $k$ increases from $-2.93 k_0$ to $1.9 k_0$ as $d$ grows from $0.02\lambda_m$ to $0.2\lambda_m$.

Then, we further realize the nonreciprocal waveguide in experiment. Despite the experimental limitations that restricted our ability to demonstrate more complex effects, the verified unidirectional propagation property serves as the foundation of our proposed unidirectional guided-wave-driven metasurfaces, providing compelling evidence for the feasibility of the proposed concept. The photographs of the fabricated nonreciprocal waveguide and the schematic of the measurement configuration are shown in Fig. 2c, here, magnetized YIG serves as the gyrotropic medium. The measured frequency-dependent S parameters of the prototype of the waveguide compared with the simulated S parameters are shown in Fig. 2d. The experimental results clearly show that there is a wide band of unidirectional propagation in the frequency range from 2.3 to 3.1 GHz (fractional bandwidth about 30%) with reverse isolation of 80 dB. The measured S parameters agree well with the simulated results with minor deviations mainly caused by the uniformity of applied static magnet in experiment. The magnetic field distribution of the electromagnet is measured and found that it has peak value at the center of the magnetic pole and decreases by ten percent at the edge. Besides, fabrication and assembling errors also deteriorate the experiment results. Despite the minor deviations, the measurement results still identify the evident unidirectional propagation effect.

Next, we investigate a periodic structure that consisting of alternating type-I waveguide of length $a$ and type-II waveguide of length $b$, as depicted in Fig. 2e. The unit cell of this structure, with a subwavelength length of $p = a + b$, supports both types of USMPs and is referred to as the USMP cell. Evidently, such structures with deep-subwavelength $d$ ($\sim \lambda/30$) can be utilized as metasurfaces for beam deflection, which extracts waves into free space by scattering two types of USMPs at their interfaces. The extracted wave has uniformly distributed phases ($\varphi_i$) spaced by $\Delta\varphi_{cl}$, which equals to the accumulated phase of USMPs traveling over a unit cell, given by

$$\Delta\varphi_{cl} = k_1 a + k_2 b + 2\delta\varphi_{12}, \qquad (4)$$

where $k_1$ is the phase constant of type-I USMP that is determined by $d$, and $k_2$ is the phase constant of type-II USMP. $\delta\varphi_{12}$ represents the phase shift from the coupling of two USMPs, which is far smaller than $\pi$ in general (see SI). The subwavelength period of the metasurface eliminates high-order diffractions, resulting in a well-defined angle of extracted beam given by $\theta = \arcsin(k_x/k_0)$, where $k_x = \Delta\varphi_{cl}/p$, also known as the Bloch wavevector. Obviously, the output angle $\theta$ can be effectively adjusted by the parameter $d$, offering a beam deflection angle range of nearly 180° based on the phase controllability of type-I USMP (see SI). On the other hand, noting that $\Delta\varphi_{cl}$ can also be tuned by adjusting parameter $a$. The $\Delta\varphi_{cl}$ was numerically calculated using full-wave finite element method (FEM), and the results are shown in Fig. 2f for $f_m = 3.587$ GHz, $f = 0.75 f_m$ and $p = 36$ mm. Here, to characterize the phase controllability of such metasurfaces, we introduce an effective index that is $\Delta\varphi_{cl}$ scaled by $k_0 p$,

$$n_{eff} = \Delta\varphi_{cl} / (k_0 p). \qquad (5)$$

As displayed in Fig. 2f, this effective index can be tuned within a range extending beyond [−1, 1], validating the phase controllability of the metasurfaces.

Let's suppose an incident wavefront in free space at $z = 0$, traveling along the positive $z$ direction. We represent the complex field across the wavefront using $U(x, 0)$ and express it as a Fourier integral

$$U(x,0) = \int_{-\infty}^{\infty} A(k_x) \exp(jk_x x) dk_x, \qquad (6)$$

where $A(k_x)$ is the spatial spectrum of the field. The consequent field $U(x, z)$ in free space ($z > 0$) can be expressed in terms of the spatial spectrum, yielding

$$U(x,z) = \int_{-k_0}^{k_0} A(k_x)e^{j(k_x x+k_z z)}dk_x + \left\{\int_{-\infty}^{-k_0} + \int_{k_0}^{+\infty}\right\} A(k_x)e^{(ik_x x-\gamma z)}dk_x, \qquad (7)$$

where $k_z = \sqrt{k_0^2 - k_x^2}$ for $|k_x| \leq k_0$ and $\gamma = \sqrt{k_x^2 - k_0^2}$ for $|k_x| > k_0$. Evidently, the field $U(x, z)$ can be divided into two parts: propagating wave with $|k_x| \leq k_0$ and evanescent wave with $|k_x| > k_0$, which are represented by the first and second terms in Eq. (7), respectively. The evanescent part attenuates rapidly along $z$, and therefore, the field pattern in far field is mainly determined by the propagating part. Thus, for achieving any desired free-space mode, we only need to construct initial wavefront $U_0(x)$ with a spatial spectrum lying within the region $|k_x| \leq k_0$, i.e.,

$$U_0(x) = \int_{-k_0}^{k_0} A(k_x)\exp(jk_x x)dk_x. \qquad (8)$$

For such a wavefront, the phase difference of the field between any two points separated by $p$ should be less than $k_0 p$, i.e., $-k_0 p \leq \arg\{U_0(x+p)\} - \arg\{U_0(x)\} \leq k_0 p$ (further discussed in the SI). For our USMP-driven metasurface, the difference of adjacent extracted phases separated by the subwavelength distance $p$ can be tuned over the range of $[-k_0 p, k_0 p]$, which corresponds to the $n_{eff}$ tune range of USMP cells larger than the interval $[-1, 1]$, so it would be suffice to implement free-space propagating wave modulation. This also greatly distinguishes from the existing metasurfaces which necessitate phase shifts cover the entire $2\pi$ phase range to achieve complete control of wavefront. In the USMP-driven metasurfaces, the scattering strength of USMPs depends on the mismatch between the modal profiles of the two USMPs. The modal spot of type-I USMP is closely related to $d$, allowing for scattering amplitude adjustment by controlling $d$. Additionally, the amplitude of the extracted wave can be significantly impacted by the size of the radiation aperture, characterized by the length $b$. As shown in Fig. 2f, for any desired phase ($n_{eff}$), $b$ can be chosen from a wide range $[0, 0.7p]$, providing ample operational freedom for amplitudes. Hence, by selecting appropriate $d$ and $a$ values, we can simultaneously achieve desired extracted phases and amplitudes. The metasurfaces that deflect beams directly from guided waves are commonly referred to as leaky-wave antennas, which typically exhibit uniform electromagnetic responses. However, through the construction of nonuniform periodic structures with USMP cells of subwavelength length $p$ having varying $d$ and $a$ values, we can achieve spatial-variant optical responses that extract and mold USMPs to any desired free-space optical modes. Therefore, the existing technology of leak-wave antennas can be regarded as a special case of the generalized formalism for USMP-driven metasurfaces.

To showcase the capability of the USMP-driven metasurfaces, we then numerically demonstrated wave focusing and Bessel-beam generation in free space directly from USMPs. For simplicity, we employed YIG material with remanence to construct metasurface cells, which can support USMPs without external magnetic field. Leveraging the unique characteristics of USMP-driven metasurfaces, we also presented numerical evidence of sub-diffraction focusing, a feat unattainable with conventional metasurfaces. Besides, by utilizing USMP cells to construct a ring cavity, vector optical vortices can be generated with quantized orbital angular momentums (OAMs). Moreover, the OAM order of radiated vortex beam can be tuned at a fixed frequency through not only adjusting the dielectric-layer thickness but also tailoring external magnetic field applied in the cavity.

**Wave focusing and bessel-beam generation**

In order to construct unidirectional guided wave-driven metasurfaces capable of complex 3D manipulation of free-space waves, 3D USMP cells are required as building blocks. By terminating 2D USMP cells in the *y* direction with a pair of subwavelength-separated metal slabs (which can be approximated as perfect electric conductors in the microwave regime), the resulting 3D USMP cells are nearly physically identical to their 2D counterparts. Leveraging these designed 3D USMP cells, it becomes possible to construct unidirectional guided wave-driven metasurfaces with desired free-space functions, as illustrated in Fig. 3a. Along the *x* axis, 3D USMP-driven metasurface is physically identical to its 2D counterpart, i.e., arbitrary $\Delta\varphi(x, y)$ along the *x* axis can be achieved through tailoring the height of the upper metal slabs $d(i, j)$ ($i, j = 1, 2, 3, \ldots$) of 3D USMP cells. To realize arbitrary 3D modulation of extracted wave, it is also necessary to have full control over the initial phase along the *y* axis, denoted as $\varphi(0, j)$. Evidently, through tailoring the heights $d(0, j)$ of the upper metal slabs of the input (type-I) waveguides to achieve different phase accumulations from the propagation of USMPs, the arbitrary initial phase $\varphi(0, j)$ of extracted wave can be achieved. At any rate, Leveraging the unique dispersion property of type-I USMPs, 3D USMP cells can offer excellent phase controllability through modulation of the height distribution of the upper metal slabs $d\{i, j\}$. This enables the extracted wave to fulfill arbitrary free-space functions, such as 3D beam directing, focusing, and vortex beam generation, as illustrated in Fig. 3a.

Let's take 3D beam focusing as an example. Through designing a USMP-driven metasurface to fulfill a 3D lens phase function $\varphi(x, y) = -k_0(\sqrt{x^2 + y^2 + F^2} - F)$, we can focus the extracted wave in free space with a designated focal length $F$. According to Eq. (4), the phase shift provided by a USMP cell between the coordinates ($x-p$) and $x$ should be

$$\Delta\varphi_{cl} = -k_0\left[\sqrt{x^2 + y^2 + F^2} - \sqrt{(x-p)^2 + y^2 + F^2}\right]. \tag{9}$$

Evidently, based on $\Delta\varphi_{cl}$, structure parameters $d\{i, j\}$ can be determined by searching the database of Fig. 2d ($a/p = 0.5$ here), and the meta-device can be achieved by spatially arranging 3D USMP cells with the designed $d\{i, j\}$. We simulated such a metalens using a full-wave FEM at 2.69 GHz. Figure 3b shows the phase distribution $\varphi(x, y)$ of extracted wave for realizing a 3D focusing with $F = 250$ mm ($\sim 2.24\lambda$) and $p = 36$ mm, the duty cycle $a/p$ of all USMP cells is fixed as 0.5. Evidently, the phase of the extracted wave satisfies a parabolic distribution. Furthermore, the simulated phase of extracted electric field agrees well with that from our theoretical calculation, along not only the horizontal direction (Fig. 3b, top), but also the vertical direction (Fig. 3b, right), further validating our USMP-driven metasurface approach. The simulated $|E_y|^2$ field distributions in both $xz$ and $yz$ planes are depicted in Fig. 3c. Obviously, unidirectional guided-wave is extracted and focused into free space by the metalens, with a focal length $F = 2.23\lambda$ in good agreement with theoretical prediction $2.24\lambda$. To check the quality of the focusing effect, we quantitatively evaluate the full width at half maximum (FWHM) of the focal spot on the focal plane, and find it is approximately $0.59\lambda$ (Fig. 3d). Such a value strongly depends on the aperture size of our metalens, and can be further reduced by enlarging the total size of our metalens. Evidently, by simply changing the distribution of $d\{i, j\}$, the focusing function of this metadevice can be switched to other functions such as beam directing and vortex beam generation. It is worth noting that, despite the phase of extracted wave from this meta-device satisfies the targeted profile, its amplitude is not uniform and actually exhibits an overall decreasing trend owing to the loss of material absorption and extraction. To achieve simultaneous control of both phase and amplitude, we should lift the constraint of constant duty cycle for all USMP cells, as done in our next design of a USMP-driven metasurface that generates a 3D Bessel beam with uniform extracted amplitude.

Bessel beams, which are solutions of the free-space Helmholtz equation, exhibit unique transverse amplitude distributions that can be described by the Bessel functions of the first kind. These beams possess remarkable properties, including non-diffraction and self-reconstruction, and can feature an extremely narrow central spot radius, on the order of one wavelength. Ideal Bessel beams carry infinite energy and are not spatially limited, their central intense region is surrounded by concentric rings that each contain an energy flux equal to that of any other. These properties make Bessel beams a promising tool for a wide range of applications in optics and photonics. As optical systems inherently possess a finite aperture, Bessel beams can only be approximated in a limited region through the superposition of multiple plane waves. This approximation can be achieved using axicons, such as conical prisms, which refract light rays symmetrically toward the optical axis. Alternatively, metasurfaces have been proposed as a means to generate Bessel beams, with these structures being referred to as meta-axicons. However, most of these meta-axicons are driven by waves in free space, which poses challenges for on-chip integration. Here, we demonstrate the generation of Bessel beams directly from photonic integrated components using USMP-driven metasurfaces.

As a proof of concept, we numerically demonstrate the generation of zero-order Bessel beam using a USMP-driven meta-axicon (Fig. 4a). Such a meta-axicon requires a spatial phase profile of the form $\varphi(x, y) = -\alpha\sqrt{x^2 + y^2}$, and the phase shift provided by a USMP cell between the coordinates $(x–p)$ and $x$ should be

$$\Delta\varphi_{cl} = -\alpha\left[\sqrt{x^2 + y^2} - \sqrt{(x-p)^2 + y^2}\right], \tag{10}$$

where $\alpha = k_0\sin\theta$ is the transverse wave number, and $\theta$ is the angle at which radiation rays cross the optical axis. To streamline the design process, we utilize L-shaped USMP cells with $p = 36$ mm to construct the meta-axicon. To be specific, this L-shaped cell can be characterized as a combination of two type-I waveguides with a cumulative length $a_i = a'_i + a''_i$ and a type-II waveguide with a length $b_i$, the length of the leading type-I waveguide remains fixed at $a'_i = p/6$ here. Because of the relatively large thickness of the dielectric layer of the latter type-I waveguide, its phase constant closely approximates $k_2$ for the type-II USMP. Hence, the phase change over the cell can be approximated as a function solely dependent on the parameter $d_i$, i.e., $\Delta\varphi_{cl} \approx k_1(d_i)p/6 + 5k_2 p/6 + 2\delta\varphi_{12}$. On the other hand, the extracted intensity can be readily regulated by manipulating the parameter $a_i$ (or $a''_i$). As a result, parameters $d_i$ and $a_i$ exert substantial influence solely over the extracted phase and intensity, respectively. The determination of structure parameters $d\{i, j\}$ can be accomplished by searching the database in Fig. 2d, taking into account the $\Delta\varphi_{cl}$ as described in Eq. (10), with a fixed ratio of $a'_i / p = 1/6$. Subsequently, the parameters $a\{i, j\}$ can be determined in a step-by-step manner by satisfying the condition of uniform radiation power of each cell, regardless of losses encountered. Consequently, all the parameters $a\{i, j\}$ and $d\{i, j\}$ can be precisely determined, and the meta-axicon can be achieved by spatially arranging such 3D L-shaped USMP cells.

The Bessel beam generated by the USMP-driven meta-axicon is numerically simulated using full-wave FEM. Figures 4b and c respectively display the simulated phase and energy of the extracted wave from each USMP cell of the meta-axicon. Evidently, simulated electric field distribution for the designed meta-axicon validates that the extracted waves have almost uniform amplitudes and spatial-variant phases as described by $\varphi(x, y) = -\alpha\sqrt{x^2 + y^2}$. Figure 4d shows the simulated Re[$E_y$] distributions for the extracted wave at both $xz$ and $yz$ planes at the designed frequency $f = 2.69$ GHz, and Fig. 4e shows the simulated intensity profile $|E_y|^2$ in $yz$ plane with $x = 0$. Both Figs. 4d and e clearly show that the extracted wave in this configuration is indeed a well-behaved Bessel beam exhibiting a clear nondiffracting feature. The propagation range of the Bessel beam is observed to be $y_{max}$ = 1204.6 mm. This maximum propagation length is close to the theoretical value using geometric optics, that is $L/(2\tan\theta) = 1236.4$ mm, where $L = 900$ mm is the length of the meta-axicon region. To characterize the performance of the generated Bessel beam, we also displayed the intensity distribution $|E_y|^2$ of the extracted wave in three $xy$ planes at different longitudinal positions ($z = 5\lambda$, $8\lambda$, and $11\lambda$). As shown in Fig. 4f, the generated transverse field patterns exhibit nice rotationally invariant symmetries with strengths that decay quickly away from the center. In Fig. 4g, we compare the intensity distribution along the $y$ axis (at $z = 8\lambda$ and $x = 0$ mm), obtained by the FEM simulation, with the zero-order Bessel function $J_0(\alpha y)$, where $\alpha = k_0\sin\theta$. Excellent agreement among these results clearly demonstrate the high quality of the Bessel beam generated by our metadevice. The full width at half maximun (FWHM) of this Bessel beam is 110.84 mm, which is close to its theoretical value, given by $\omega_{FWHM} = 2.25/\alpha = 116.76$ mm.

In addition, it is important to evaluate the efficiency of the proposed metasurfaces. In the simulation, the absorption losses of remanence materials are characterized by the damping coefficient $v = 10^{-3}\omega$ [39]. The simulated utilization efficiencies (UEs) of proposed metasurfaces are 70% for wave focusing and 13% for Bessel-beam generation, which is calculated as the ratio between phase-modulated output and the energy decrease of USMPs inside the waveguide. It should be noted that the lower UEs observed in Bessel beam excitation can be attributed to the relatively larger thickness of the air layer within the latter type-I waveguide, resulting in a smaller amount of extracted energy. In fact, the UEs of Bessel-beam generation can also exceed 70% after parameter optimization (further discussed in the SI). The reduced efficiency primarily stems from material absorption losses.

**Sub-diffraction focusing**

By utilizing USMP-driven metasurfaces composed of deep-subwavelength cells, we have successfully constructed a metasurface capable of manipulating both propagating and evanescent wave components, thereby achieving sub-wavelength focusing in the near-field regime and ultimately breaking the diffraction limit. Subwavelength focusing holds great potential for improving the resolution of imaging systems, enhancing detector sensitivity, and expanding a multitude of applications, including but not limited to superresolution microscopy, nanoscopy, nanolithography, and optical trapping. The key to achieving subwavelength resolution and overcoming the diffraction limit lies in the precise tailoring of the evanescent spectrum of an aperture field distribution.

In 1968, Victor Veselago first discussed the electromagnetic properties of materials exhibiting simultaneously negative permittivity and permeability and proposed focusing devices that operate based on negative refraction[40]. Notably, John Pendry showed in 2000 that a planar slab composed of a negative refractive index material can manipulate near fields to achieve perfect imaging, theoretically realizing a complete reconstruction of both near and far-field sources[2]. Subsequently, negative refractive index and negative permittivity superlenses were experimentally verified[41-43]. In addition, researchers introduced a related but alternative approach in 2007 that relies on patterned, grating-like surfaces to achieve subwavelength resolutions[44-46]. However, these structures are limited to object placements in close proximity to the focusing plate (much smaller than the wavelength). An alternative solution using superoscillations has been proposed to overcome this limitation, yet it comes with the expense of significant scattering into parasitic propagating modes[47-49]. It is essential to acknowledge that all known methods for subwavelength field concentration and focusing have their specific limitations and drawbacks, and thus it is crucial to explore additional possibilities that may offer complementary advantages and further enhance our comprehension of these phenomena.

It is worth noting that, although the initial wavefront $U_0(x)$ in the above USMP-driven metasurfaces are constructed by propagating waves with $|k_x| \leq k_0$, it can also be constructed beyond the spatial spectrum region $|k_x| \leq k_0$, i.e. evanescent waves can also be incorporated. Furthermore, while the length of USMP cells in the aforementioned examples is set as $p \sim \lambda/3.1$, it is completely feasible to reduce the cell sizes to a deep-subwavelength level. With these unique characteristics, the USMP-driven metasurface presents a promising approach to realizing near-field subwavelength focusing that cannot be achieved with conventional metasurfaces. Full-wave simulations demonstrate that it is able to break the diffraction limit and provide near-field focusing with subwavelength hotspot size.

We numerically demonstrate the generation of near-field subwavelength focusing using a USMP-driven metadevice consists of L-shaped USMP cells (Fig. 5a). Such a metadevice requires an initial wavefront $U_0(x)$ of the form

$$U_0(x) = \int_{-k_0}^{k_0} \cos(kx) e^{-jf_0\sqrt{k_0^2 - k^2}} dk + \left( \int_{-k_{max}}^{-k_0} + \int_{k_0}^{k_{max}} \right) \cos(kx) e^{f_0\sqrt{k^2 - k_0^2}} dk \ , \tag{11}$$

where $k_{max}$ determines the hotspot size $d_{FWHM} = 2.783/k_{max}$. The case of $k_{max} = k_0$ corresponds to the usual diffraction-limited focusing without the contribution of evanescent modes. To achieve sub-diffraction focusing, $k_{max}$ must be greater than $k_0$. We consider $k_{max} = 2.6k_0$ here, which corresponds to a hotspot size of $d_{FWHM} = 0.17\lambda_0$, and the black lines in Fig. 5b illustrate the spatial amplitude and phase profiles described by Eq. (11). The large amplitude and phase fluctuations within a single wavelength range pose a formidable challenge for conventional metasurfaces in attaining near-field subwavelength focusing. This predicament is primarily attributed to the resonant nature of meta-atoms, which makes it difficult to manipulate wavefronts on deep subwavelength scales.

In our design, we adopt L-shaped USMP cells with a deep subwavelength size $p = \lambda/20$ to construct the metadevice. The design scheme is similar with the aforementioned meta-axicon. Figure 5e shows the electric field intensity distribution $|E_y|^2$ of the extracted wave from the designed metadevice. Evidently, subwavelength focusing is achieved due to the excitation of evanescent modes. In Fig. 5c we plot the $|E_y|^2$ distribution along a line crossing the focal point (i.e., at $z = 0.1\lambda$), and the FWHM is calculated as $0.17\lambda$, much smaller than the case when only propagation modes are contributing ($0.443\lambda$), thereby overcoming the diffraction limit. The simulated electric-field distribution on the designed metadevice is displayed in Fig. 5b, further validating that the extracted waves have targeted spatial amplitude and phase profiles described by Eq. (11) (black lines in Fig. 5b). For comparative purposes, the theoretically computed field pattern ($|E_y|^2$) excited by an incident wavefront $E_y = U_0(x)$, with $U_0(x)$ described by Eq. (11), is also plotted in Fig. 5d. Evidently, the simulated electric-field pattern $|E_y|^2$ radiated from the USMP-driven metadevice in Fig. 5e agrees well with the targeted one of Fig. 5d. Figures 5f and 5g show simulated magnetic-field profiles of the nonzero field components $|H_x|^2$ and $|H_z|^2$ at the designed frequency $f = 2.69$ GHz. As before, the field pattern of $|H_x|^2$ is almost identical to that of $|E_y|^2$.

**Tunable ring-cavity OAM source**

Utilizing 3D USMP cells to construct a ring cavity, we are able to create a tunable source that directly emits OAM beams. The OAM beams are helically phased beams comprising an azimuthal phase term $\exp(il\theta)$, and they carry an OAM of $l\hbar$ per photon, where $l$ is an integer known as the topological charge, and $\theta$ is the azimuthal angle with respect to the propagation direction. Different from spin angular momentum (SAM), which is associated with photon spin and manifested as circular polarization, the OAM is linked to the spatial distribution with vortex nature. In contrast to the SAM, which can only take two values of $\pm\hbar$, the OAM is unbounded. Thus, EM waves can have an infinite number of orthogonal OAM states theoretically. Due to this unique property, OAM beams are considered as potential candidates for encoding information in both classical and quantum optical systems. A conventional system for generating OAM beams is usually based on the combination of a bulk source with additional phase-front shaping components, such as spatial phase plates, spatial light modulations, and metasurfaces. However, relying on rather different device technologies and material platforms, this approach is not easily scalable and integratable. By using USMP-driven metasurface technology, this fundamental limitation can be overcome, and compact, integratable and tunable ring-cavity OAM sources can be realized.

A ring cavity generally supports degenerate whispering gallery modes (WGMs): clockwise (CW) and counterclockwise (CCW) propagating modes[28]. These modes circulate inside the cavity and carry large OAM. Because of the inversion symmetry of the cavity, the OAMs of CW and CCW eign-WGMs have opposite signs. These modes can be excited simultaneously, and consequently their carried OAMs cancel each other. Controllable OAM emission has been demonstrated by implementing a phase-gradient metasurface integrated on a ring cavity. This metasurface breaks the degeneracy of two counter-propagating WGMs so that only one WGM can couple to the free space emission. Here, using USMP cells to construct a ring cavity, we develop a straightforward approach to achieve controllable OAM emission. In such cavity, the inversion symmetry is intrinsically broken, and only one of either CW or CCW WGM is allowed to propagate, which is determined by the magnetization direction in the YIG materials. In contrast to the previous approach using phase-gradient metasurface, only unidirectional power circulation can occur in our cavity, which eliminates the undesired spatial hole-burning effect resulted from the interference pattern of two counter-propagating WGMs. So the targeted WGM in our ring cavity unidirectionally circulates carrying OAM through the azimuthally continuous phase evolution.

We consider a ring cavity that consists of $N$ uniform USMP cells periodically ranged along the azimuthal direction (Fig. 6a). Suppose USMPs in the cavity travel along the CCW direction. The eigen-WGM in the cavity has the feature of azimuthal resonance and, for the $M$th-order mode, the following condition should be satisfied

$$N\Delta\varphi_{cl} = n_{eff} k_0 Np = 2\pi M, \qquad (12)$$

where $n_{eff}$ is the effective index of the USMP cell and $p$ is the period length. In this cavity, the USMPs are coherently scattered at the locations $\theta_n = 2\pi n/N$, where $n \in [0, N-1]$, resulting in the extracted phases $\varphi_n = 2\pi nM/N$. The extracted phases increase linearly from 0 to $2\pi M$ along the ring perimeter, thereby creating a vortex beam with topological charge $l = M$. As $n_{eff}$ can be continuously tuned over the range $[-1, 1]$ by varying frequency, the OAM order of radiated beam can vary with frequency. More importantly, $n_{eff}$ can be tuned at a fixed frequency through varying external magnetic field applied in the YIG material, so we can achieve OAM sources with tunable topological charges. Note that only USMPs with $|n_{eff}| < 1$ can be coupled to the free space emission, which gives a limitation to the range of achievable OAM states.

Let the quantity $l = N\Delta\varphi_{cl}/2\pi$ numerically characterize the phase shift of USMPs traveling over $N$ USMP cells. This quantity corresponds to the OAM order of eigen WGM in a ring cavity that consists of $N$ USMP cells. The quantity $l$ can be calculated using the full-wave FEM simulation for the USMP cell with periodic boundary condition. In the presence of external magnetic field ($H_0$), the tensor components in Eq. (1) become $\mu_1 = 1 - \omega_0\omega_m/(\omega^2-\omega_0^2)$ and $\mu_2 = \omega_m\omega/(\omega^2-\omega_0^2)$, where $\omega_0 = 2\pi\gamma H_0$ ($\gamma$ is the gyromagnetic ratio) and $\omega_m$ is the characteristic circular frequency. Figure 6b shows the $l$ value as a function of the external magnetic field and frequency ($f$). Here, we assume $N = 86$, and take the parameters for the USMP cell as follows: $\omega_m = 10\pi \times 10^9$ rad/s ($f_m = 5$ GHz), $p = 3$ mm, $a = 1$ mm and $d = 1.212$ mm (the dispersion relations of USMPs for type-I and type-II waveguides are displayed in the SI). As the type-I USMP in the cell only exists in the range from $\omega_0+0.5\omega_m$ to $\omega_0+\omega_m$, where $\omega_0 = 2\pi\gamma H_0$ ($\gamma$ is the gyromagnetic ratio), the frequency region for unidirectional WGMs in the ring cavity varies with $H_0$. In the colormap, some $l$ integers are marked with stars for the frequency $f = 8.5$ GHz, and they are found to be really the OAM orders of WGMs in the corresponding cavity. The phase shift for a single USMP cell is $\Delta\varphi_{cl} = n_{eff}k_0p$, and $n_{eff}$ can be tuned by varying $H_0$ over the range $[-1, 1]$. As a result, the OAM order has at least a tunable range of $[-l_{max}, l_{max}]$, where $l_{max} = \text{int}[Np/\lambda_0]$ with $\lambda_0$ being the designed wavelength. Clearly, the radiated OAM can be adjusted at a fixed frequency through varying external magnetic field $H_0$. In addition, the $n_{eff}$ value varies with frequency, and so does the OAM order $l$ of WGM in the ring cavity. Thus, our ring cavity can output vortex beams with different OAM orders for different frequencies, as indicated in Fig. 6c, where $H_0 = 1785$ Gs. For $d = 1.212$ mm, the OAM order $l$ can vary from −7 to 7 over the frequency range [8.15, 9.2 ] GHz (see SI). Besides, it is also available for our ring source to output vortex beams with the same OAM order over a frequency range by properly adjusting the external magnetic field. Therefore, based on the nonresonant mechanism of the metasurface, our ring OAM source can operate over a wide frequency range.

To verify the controllability of the OAM emission with the external magnetic field $H_0$, as marked in Fig. 6b, we displayed the simulated results of OAM radiation from the ring cavity with three representative magnetic field $H_0 = 1.92$, 1.86, and 1.57 kGs in Fig. 6d1-f3 at the designed frequency $f = 8.5$ GHz (complete results are displayed in the SI). The radiated OAM beams is radially polarized. The electric field $E_r$ has a spiral pattern, and its phase changes by $-7 \times 2\pi$, $-4 \times 2\pi$ and $5 \times 2\pi$ respectively upon one full circle around the center of the vortex, indicating the topological charge $l = -7, -4$ and 5, matching our simulation result in Fig. 6b. Moreover, the intensity of the electric fields are spatially distributed in a doughnut shape with a dark core in the center, which are due to the topological phase singularity at the beam axis, where the phase becomes discontinuous. On the other hand, according to Fig. 6c, the OAM charge can also be controlled by varying frequency $f$. To illustrate this, Fig. 6g1-i3 shows three OAM states of $l = -6, 2$ and 7 at frequencies $f = 8.2, 8.74$, and 9.2 GHz, which agree well with those from our theoretical calculation in Fig. 6c. Furthermore, Fig. 6c indicates that the OAM charge can also be tailored by varying structure parameter $d$ (see SI).

**Discussion**

Our USMP-driven metasurfaces, consisting of uniform or nonuniform USMP cells of subwavelength length, provide a highly versatile and compact platform for achieving free-space optical functionalities directly from surface plamonic waves. In the microwave domain, we have experimentally verified the existence of such USMPs in a metal-air-gyromagnetic structure. Different from existing metasurfaces with meta-atoms, which induce abrupt phase shifts in the optical path, the present metasurfaces rely on gradual phase accumulation from the propagation of USMPs to mold optical wavefront in subwavelength scale. Such metasurfaces can provide unparalleled phase controllability to free-space wave propagation based on the unique dispersion property of USMP. We have demonstrated the generation of a Bessel beam and wave focusing using these structures at microwave frequencies. In addition, by taking advantage of the unidirectional propagation of USMPs, we used a ring-cavity formed by bi-USMP cells to generate optical vortices that have a helical wavefront and carry OAM with a designable order. Besides, the dispersion property of USMPs can be flexibly tailored by either geometric parameter or external magnetic field, thus facilitating the realization of dynamic controllability of USMP-driven metasurfaces. We have also demonstrated the tunability of OAM order of radiation with the geometric parameter and magnetic field at a designed frequency. The design strategies presented in the paper allow one to tailor in an almost arbitrary way the phases and amplitudes of an optical wavefront. On the basis of the demonstrated design principle, more complex functionalities can be realized, such as USMP-driven holograms, virtual reality displays, and so forth. Furthermore, as our metasurfaces can control wavefronts at deep subwavelength scales, they are capable to realize desired fine field patterns in near field. Using these structures, we have demonstrated the generation of near-field subwavelength focusing that is difficult to achieve with conventional metasurfaces. Such near-field manipulation can break the optical diffraction limit, thus yielding important applications in optical detection, optical sensing and high-resolution optical imaging. The developed technology can be extended to higher frequency regimes and could have major implications for integrated optics[5,50].

In order to extend the concept of USMP-driven metasurfaces to high-frequency regimes, achieving robust USMPs with complete phase controllability is key. USMP relies heavily on nonreciprocal materials with a strong magneto-optical effect to break time-reversal symmetry in the guiding system. Magneto-optical materials are divided into two types: gyromagnetic and gyroelectric. In the microwave regime, ferrites are gyromagnetic material that can support TE-polarized USMPs. In higher frequency regimes, such as terahertz and visible frequencies, commonly considered magneto-optical materials are gyroelectric. In the terahertz regime, under an external magnetic field of normal strength, semiconductors exhibit strong gyroelectric anisotropy. It has been confirmed both theoretically and experimentally that the semiconductor-opaque medium interface can support TM-polarized USMPs at terahertz frequencies[51,52]. These USMPs are robust against disorders or even nonlocal material effects based on semiconductor materials' bandgap with topologically nontrivial properties. Additionally, a semiconductor-medium-metal structure can also support terahertz USMPs when the medium layer is sufficiently thin, but these terahertz USMPs normally have limited phase controllability because their dispersion curves can only cover half of the air light cone. Nonetheless, our theoretical calculations show that a magnetically opaque material-semiconductor-perfectly magnetic conductor structure *can* support *terahertz* robust USMPs with complete phase controllability (see SI). We note from the above that unidirectional modes are advantageous for obtaining unique dispersion properties, and it is thus prudent designing suitable guiding structures with periodically structured interfaces that can support USMPs with judicious dispersion. Evidently, from the above discussion and Fig. S9 of the SI, USMP-driven metasurfaces *in the terahertz regime* are indeed feasible.

Concerning the optical regime, we note that normally it is challenging to achieve unidirectional modes, owing to the weak gyroelectric effect in magneto-optical materials under normal magnetic field strength. However, as has recently been shown theoretically by Kempa *et al*. (Ref. [51]), by substantially reducing the *diagonal* elements of the permittivity tensor in layered plasmonic-gyroelectric metamaterials, one can significantly enhance the gyroelectric effect in the optical regime. Such a gyroelectric metamaterial was used in Ref. [51] as the background material to construct a photonic crystal, and unidirectional photonic-crystal edge mode was achieved in the optical regime[51]. Building on that work, and for our objectives herein, one may thus envision designing gyroelectric metamaterials at optical frequencies with bulk-mode bandgaps possessing nontrivial topological properties, which are characterized by nonzero gap Chern numbers[52]. With the use of such gyroelectric metamaterials, USMPs can be realized in optical regime. To demonstrate this, we have developed a gyroelectric metamaterial composed of alternating magneto-optical material ($CuFe_2O_4$)[53] and plasmonic material (Au) layers, resulting in a topologically nontrivial bandgap at visible frequencies. As shown in the SI, visible USMPs can be supported by this metamaterial when interfaced with an opaque medium, and they exist for small wavenumbers in the entire bandgap. Although the phase controllability of USMPs in this simple configuration is limited, it is entirely possible that (visible) USMPs in complex plasmonic structures possess complete phase controllability, as their dispersion properties are closely dependent on the material configuration. While the optical regime is not the focus of our work, the example presented here demonstrates the feasibility of USMPs in this regime. Therefore, it can be concluded that USMP-driven metasurfaces should be feasible in the optical regime as well.

**Methods**

**Numerical simulations**

The theoretical data presented in Figs. 2b and 2f were obtained using the software of MATLAB. All numerical simulations were conducted using the finite element method (FEM), with third-order finite elements and at least 10 mesh steps per wavelength to ensure the accuracy of the calculated results. In both 2D and 3D simulations, unidirectional surface magnetoplasmons (USMPs) were excited by point sources and line currents, respectively, placed in a unidirectional waveguide that was connected to the metasurface system. The effects of material absorption of yttrium iron garnet (YIG) were considered in the damping coefficient, which was set to $\nu = 10^{-3}\omega$, where $\omega$ is the operating angular frequency.

To generate free-space wave manipulation with desired phase profile, we first simulated a periodic structure consisting of uniform USMP cells (Fig. 2e) using the FEM. The phase difference between the extracted waves from two neighboring cells was calculated as $\Delta\varphi$. We then systematically varied the geometrical parameters of the USMP cells to obtain phase map (Fig. 2f). By analyzing the desired phase profiles, we obtained the target phase difference distribution $\Delta\phi(x, y)$. Subsequently, we identified the structural parameters $d\{i, j\}$ that correspond to the target phase distribution by searching the phase map. This approach allowed us to accurately determine the detailed structure of the USMP cells required for achieving the desired phase distribution.

Finally, we describe the approach for achieving precise manipulation of free-space waves, encompassing both desired phases and amplitudes profiles. As mentioned in the main text, firstly, the determination of structure parameters $d\{i, j\}$ can be accomplished by searching the database presented in Fig. 2f. Subsequently, the parameters $a\{i, j\}$ can be determined in a step-by-step manner by satisfying the desired radiation amplitude of each cell. This approach enables us to accurately determine the detailed structure of the USMP cells for free-space waves manipulation with not only the desired phase but also the desired amplitude distributions.

**Experimental setup**

In experiment, a YIG slab that is 40 mm long is placed in metallic waveguide between two ceramics, with an air gap of $d = 2$ mm from one of the sidewalls. The waveguide is fed with two SMA ports, filled with high dielectric constant ceramic to lower the cutoff frequency, which can support TE mode wave in our expected frequency band. The YIG has the material parameters of $4\pi M_s = 700$ Gs, $\varepsilon_m = 13.5$ and $\Delta H = 90$ Oe. A strip of ridge is carved out in each ceramic to obtain better port match, the two ports are placed in opposite directions so that only one kind of ceramic structure is needed with flip arrangement. An electromagnet provides the static magnet bias with strength $H_0 = 400$ Oe.

**Acknowledgments**

This work was supported by National Natural Science Foundation of China (Nos. 12104401, 62075197, 62101496). K.L.T. was supported by the General Secretariat for Research and Technology (GSRT) and the Hellenic Foundation for Research and Innovation (HFRI) under Grant No. 4509.

**Author contributions**

S.L. conceived the concept of unidirectional guided-wave-driven metasurfaces, and together with L.S. performed the initial analysis and simulations. T. J. fabricated the devices, carried out the measurements, and analyzed the experimental data; All authors contributed to the final aspects of the theory, simulations, and discussions. L.S. wrote a first draft of the paper, which was then finalized by input from S.L. and K.L.T.. L.S., and K.L.T. supervised the work.

**Competing interests**

The authors declare no competing interests.

# References

1. Barnes, W. L., Dereux, A. & Ebbesen, T. W. Surface plasmon subwavelength optics. *Nature* **424**, 824–830 (2003).

2. Pendry, J. B. Negative Refraction Makes a Perfect Lens. *Phys. Rev. Lett.* **85**, 3966–3969 (2000).

3. Kauranen, M. & Zayats, A. V. Nonlinear plasmonics. *Nat. Photon.* **6**, 737–748 (2012).

4. Bozhevolnyi, S. I., Volkov, V. S., Devaux, E., Laluet, J.-Y. & Ebbesen, T. W. Channel plasmon subwavelength waveguide components including interferometers and ring resonators. *Nature* **440**, 508–511(2006).

5. Bi, L. et al. On-chip optical isolation in monolithically integrated non-reciprocal optical resonators. *Nat. Photonics* **5**, 758–762 (2011).

6. Brion, J. J., Wallis, R. F., Hartstein, A. & Burstein, E. Theory of Surface Magnetoplasmons in Semiconductors. *Phys. Rev. Lett.* **28**, 1455–1459 (1972).

7. Yu, Z., Veronis, G., Wang, Z. & Fan, S. H. One-way electromagnetic waveguide formed at the interface between a plasmonic metal under a static magnetic field and a photonic crystal. *Phys. Rev. Lett.* **100**, 023902 (2008).

8. Shen, L. F., You, Y., Wang, Z. Y. & Deng, X. H. Backscattering-immune one-way surface magnetoplasmons at terahertz frequencies. *Opt. Express* **23**, 950–962 (2015).

9. Zhu, H. & Jiang, C. Broadband unidirectional electromagnetic mode at interface of anti-parallel magnetized media. *Opt. Express* **18**, 6914–6921 (2010).

10. Hu, B., Wang, Q. J. & Zhang, Y. Broadly tunable one-way terahertz plasmonic waveguide based on nonreciprocal surface magneto plasmons. *Opt. Lett.* **37**, 1895–1897 (2012).

11. Kuzmiak, V., Eyderman, S. & Vanwolleghem, M. Controlling surface plasmon polaritons by a static and/or time-dependent external magnetic field. *Phys. Rev. B* **86**, 045403 (2012).

12. Gangaraj, S. A. H. & Monticone, F. Do truly unidirectional surface plasmon-polaritons exist?. *Optica* **6**, 1158–1165 (2019).

13. Liang, Y. et al. Tunable unidirectional surface plasmon polaritons at the interface between

14. Buddhiraju, S. et al. Absence of unidirectionally propagating surface plasmon-polaritons at nonreciprocal metal-dielectric interfaces. *Nat. Commun.* **414**, 674 (2020).

15. Lu, L., Joannopoulos, J. D. & Soljacic, M. Topological photonics. *Nat. Photonics* **8**, 821–829 (2014).

16. Hartstein, A., Burstein, E., Maradudin, A. A., Brewer, R. & Wallis, R. F. Surface polaritons on semi-infinite gyromagnetic media. *J. Phys. C.: Solid State Phys.* **6**, 1266–1276 (1973).

17. Zhang, X., Li, W. & Jiang, X. Confined one-way mode at magnetic domain wall for broadband high-efficiency one-way waveguide, splitter and bender. *Appl. Phys. Lett.* **100**, 041108 (2012).

18. Yu, Z. H., Wang, Z. Y., Shen, L. F. & Deng, X. H. One-way electromagnetic mode at the surface of a magnetized gyromagnetic medium. *Electron. Mater. Lett.* **10**, 969–973 (2014).

19. Tsakmakidis, K. L. et al. Breaking Lorentz reciprocity to overcome the time-bandwidth limit in physics and engineering. *Science* **356**, 1260–1264 (2017).

20. Yu, N. F. et al. Light propagation with phase discontinuities: generalized laws of reflection and refraction. *Science* **334**, 333–337 (2011).

21. Sun, S. L. et al. Gradient-index meta-surfaces as a bridge linking propagating waves and surface waves. *Nat. Mater.* **11**, 426–431 (2012).

22. Ni, X. J., Emani, N. K., Kildishev, A. V., Boltasseva, A. & Shalaev, V. M. Broadband light bending with plasmonic nanoantennas. *Science* **335**, 427(2012).

23. Sun, S. et al. High-efficiency broadband anomalous reflection by gradient meta-surfaces. *Nano Lett.* **12**, 6223–6229 (2012).

24. Grady, N. K. et al. Terahertz metamaterials for linear polarization conversion and anomalous refraction. *Science* **340**, 1304–1307 (2013).

25. Li, X. et al. Flat metasurfaces to focus electromagnetic waves in reflection geometry. *Opt. Lett.* **37**, 4940–4942 (2012).

26. Chen, X. Z. et al. Dual-polarity plasmonic metalens for visible light. *Nat. Commun.* **3**, 1198 (2012).

27. Lin, D. M., Fan, P. Y., Hasman, E. & Brongersma, M. L. Dielectric gradient metasurface optical elements. Science **345**, 298–302 (2016).

28. Miao, P. et al. Orbital angular momentum microlaser. *Science* **353**, 464–467 (2016).

29. Zheng, G. X. et al. Metasurface holograms reaching 80% efficiency. *Nat. Nanotechnol.* **10**, 308–312 (2015).

30. Guo, X. X., Ding, Y. M., Chen, X., Duan, Y. & Ni, X. J. Molding free-space light with guided wave-driven metasurfaces. *Sci. Adv.* **6**, eabb4142 (2020).

31. Li, Z. Y. et al. Controlling propagation and coupling of waveguide modes using phase-gradient metasurfaces. *Nat. Nanotechnol.* **12**, 675–683 (2017).

32. Meng, Y. et al. Optical meta-waveguides for integrated photonics and beyond. *Light: Sci. Appl.* **10**, 235 (2021).

33. Lezec, H. J. et al. Beaming light from a subwavelength Aperture. *Science* **297**, 820–822 (2002).

34. Yu, N. F. et al. Small-divergence semiconductor lasers by plasmonic collimation. *Nat. Photonics* **2**, 564–570 (2008).

35. Dolev, I., Epstein, I. & Arie, A. Surface-plasmon holographic beam shaping. *Phys. Rev. Lett.* **109**, 203903 (2012).

36. Ozaki, M., Kato, J. I. & Kawata, S. Surface-plasmon holography with white-light illumination. *Science* **332**, 218–220 (2011).

37. Chen, Y. H., Huang, L., Gan, L. & Li, Z. Y. Wavefront shaping of infrared light through a subwavelength hole. *Light. Sci. Appl.* **1**, e26 (2012).

38. Chen, J., Li, T., Wang, S. M. & Zhu, S. N. Multiplexed holograms by surface plasmon propagation and polarized scattering. *Nano Lett.* **17**, 5051–5055 (2017).

39. Pozar, D. M. Microwave Engineering, *John Wiley & Sons*, New York (2011).

40. Veselago, V. G. The electrodynamics of substances with simultaneously negative values of permittivity and permeability. *Sov. Phys. Usp.* **10**, 509 (1968).

41. Grbic, A. & Eleftheriades, G. V. Overcoming the Diffraction Limit with a Planar Left-Handed Transmission-Line Lens. *Phys. Rev. Lett.* **92**, 117403 (2004).

42. Taubner, T., Korobkin, D., Urzhumov, Y., Shvets, G. & Hillenbrand, R. Near-field microscopy through a sic superlens. *Science* **313**, 1595 (2006).

43. Fang, N., Lee, H., Sun, C. & Zhang, X. Sub-diffractionlimited optical imaging with a silver superlens. *Science* **308**, 534 (2005).

44. Marks, D. & Carney, P. S. Near-field diffractive elements. *Opt. Lett.* **30**, 1870 (2005).

45. Merlin, R. Radiationless electromagnetic interference: Evanescent-field lenses and perfect focusing. *Science* **317**, 927 (2007).

46. Grbic, A., Jiang, L. & Merlin, R. Near-field plates: Subdiffraction focusing with patterned surfaces. *Science* **320**, 511 (2008).

47. Huang, F. M. & Zheludev, N. I. Super-resolution without evanescent waves. *Nano Lett.* **9**, 1249 (2009).

48. Rogers, E. et al. A super-oscillatory lens optical microscope for subwavelength imaging. *Nat. Mater.* **25**, 432 (2012).

49. Yuan, G. H., Rogers, E. T. & Zheludev, N. I. Achromatic super-oscillatory lenses with sub-wavelength focusing. *Light: Sci. Appl.* **6**, e17036 (2017).

50. Gardes F. Y. et al. Micrometer size polarisation independent depletion-type photonic modulator in Silicon On Insulator. *Opt. Express* **15**, 5879–5884 (2007).

51. Wu, X., Ye, F., Merlo, J. M., Naughton, M. J. & Kempa, K. Topologically protected photonic edge states in the visible in plasmo-gyroelectric metamaterials. *Adv. Opt. Mater.* **6**, 1800119 (2018).

52. Silveirinha, M. G. Chern invariants for continuous media. *Phys. Rev. B* **92**, 125153 (2015).

53. Veis, M. et al. Complete permittivity tensor in sputtered CuFe2O4 thin films at photon energies between 2 and 5 ev. *Materials* **6**, 4096–4108 (2013).

# Figures

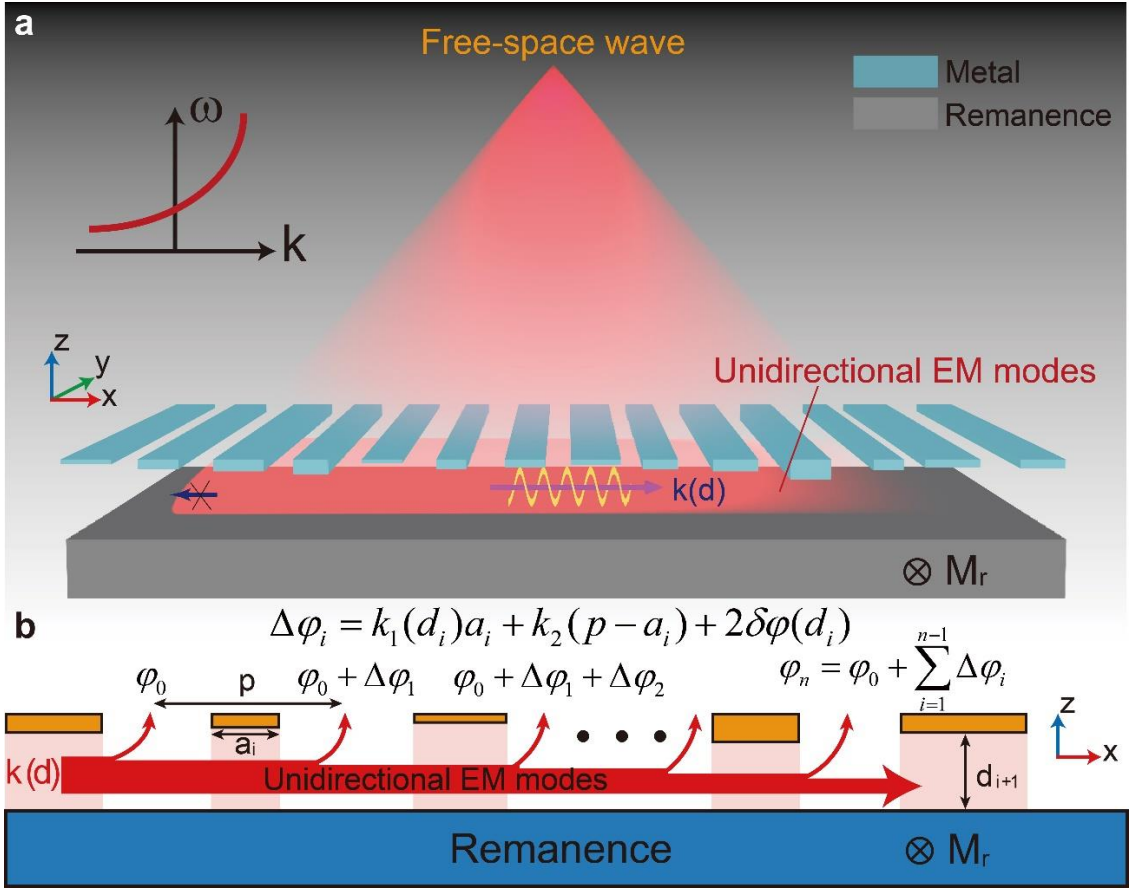

**Fig. 1| Working principle of unidirectional guided-wave-driven metasurfaces. a** Schematic of a unidirectional guided-wave-driven metasurface. The phase and intensity of the extracted wave from unidirectional waveguide by each cell can be tuned individually. An array of cells work collaboratively to form certain wavefronts and fulfill different functions, such as beam focusing. **b** Illustration of the wavefront formation of the extracted wave. The phase difference of the extracted waves from two neighboring cells is contributed from two parts: the phase accumulation $k_1(d_i)a_i + k_2(p-a_i)$ from the unidirectional guided-wave propagation and the phase shift $2\delta\varphi(d_i)$ from the coupling of two unidirectional guided modes with each other in each cell. As a result, the phase difference of the extracted waves from two neighboring cells can be expressed as $\Delta\varphi_i = k_1(d_i)a_i + k_2(p-a_i) + 2\delta\varphi(d_i)$. The superposition of phase difference $\Delta\varphi_i$ constitute the phase of extracted wave from the *n*-th cell $\varphi_n = \varphi_0 + \sum_{i=1}^{n-1}\Delta\varphi_i$, where $\varphi_0$ is the initial phase of the incidence.

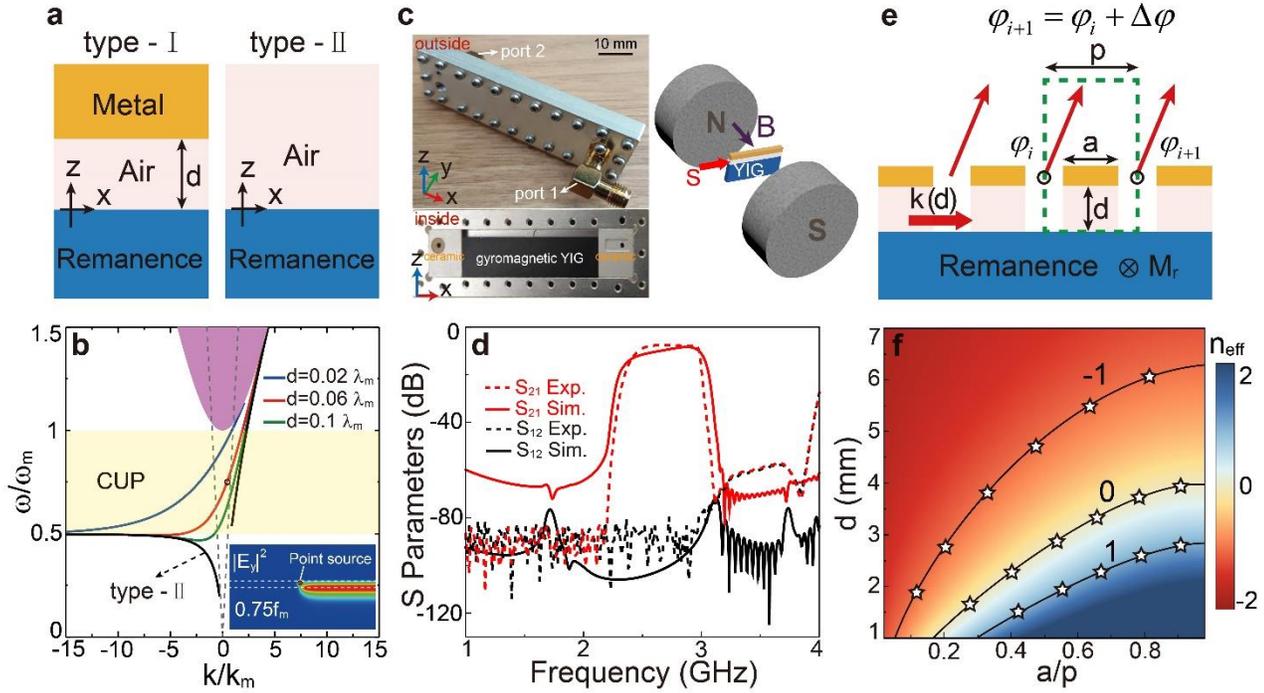

**Fig. 2 | Phase controllability of the unidirectional guided-wave. a** Schematic diagram of waveguides supporting type-I and type-II USMP at microwave frequencies, and **b** the corresponding dispersion relations of USMPs for various thicknesses of the dielectric layer. The shaded rectangular area indicates the one-way region for the waveguide, and the other shaded areas indicate the zones of bulk modes in the gyromagnetic materials ($\varepsilon_m = 15$). Inset: simulated electric-field intensity distribution excited by a point source placed in the middle of waveguide structure with $d = 0.06\lambda_m$ and frequency $\omega = 0.75\omega_m$. **c** Left: Photograph of the fabricated YIG-dielectric-metal unidirectional waveguide. Right: schematic of the measurement configuration. **d** Simulated (solid) and measured (dashed) S-parameter of the unidirectional waveguide displayed in (**c**). **e** Schematic picture of the uniform metasurface formed by alternating type-I and type-II waveguides shown in (**a**). The period length, duty cycle, and dielectric-layer thickness are denoted by $p$, $a/p$, and $d$ respectively. The phases of extracted wave from two neighboring cells are $\varphi_i$ and $\varphi_{i+1}$ with a difference $\Delta\varphi$. **f** Pseudocolor map of simulated $n_{eff} = \Delta\varphi_{cl}/(k_0 p)$ for the structure of (**e**) in a parameter space spanned by dielectric thickness ($d$) and duty cycle ($a/p$) at 2.69 GHz. The three black lines indicate $n_{eff}$ covering [−1, 1] range with an even interval. The black stars indicate the theoretically analyzed results from Eq. (4).

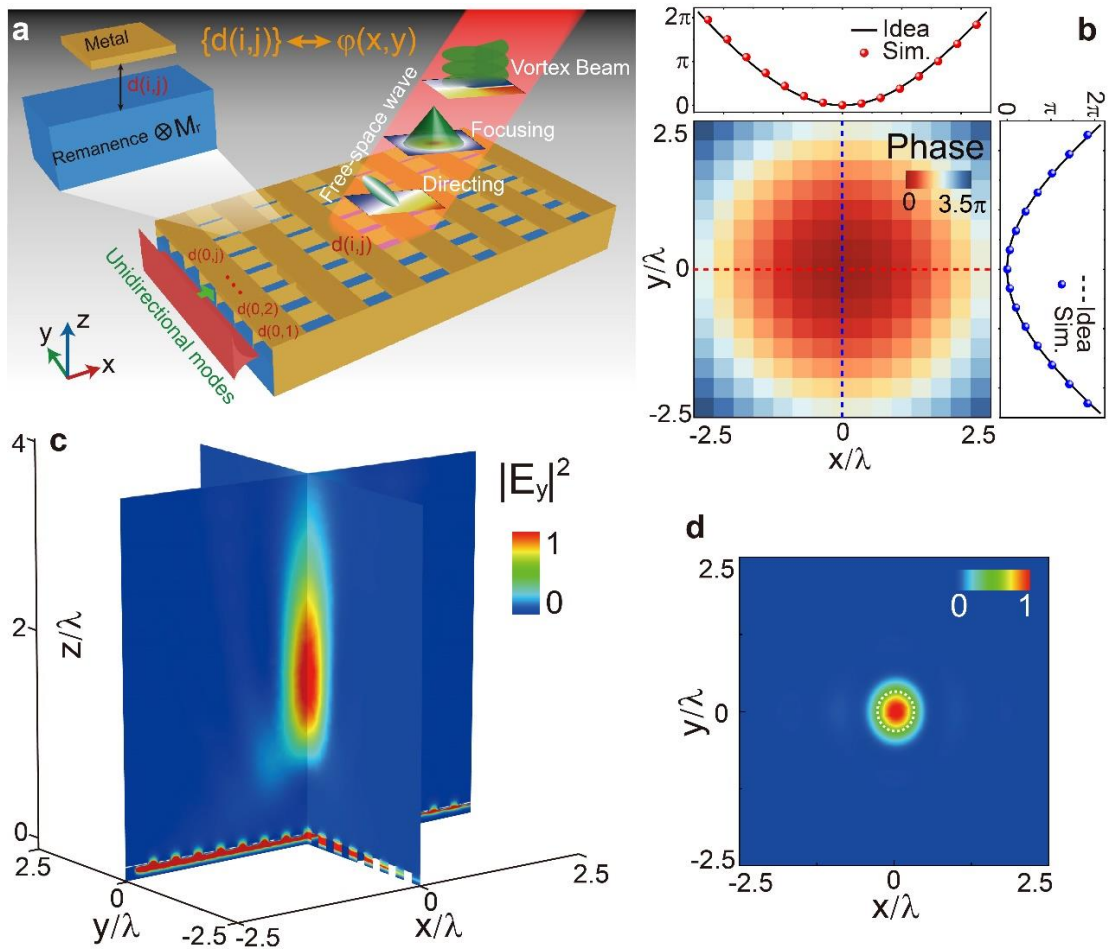

**Fig. 3| Demonstration of 3D manipulation of free-space wave with a unidirectional guided wave-driven metasurface. a** Schematic of the designed meta-device which is composed of 2D USMP cells terminated in the *y* direction with a pair of metal slabs separated by a subwavelength distance. Through modulate the height distribution of the upper metal slabs $d\{i, j\}$, extracted wave from the unidirectional waveguide can fulfill arbitrary functions, such as free-space beam directing, focusing and vortex beam generation. **b** The simulated phase distribution of extracted wave from each USMP cells of the designed meta-device, together with its corresponding horizontal and vertical cuts. The target phase profiles along both the horizontal and vertical directions are also presented for comparison. **c** Simulated 3D electric field intensity $|E_y|^2$ distribution of the designed meta-device, the extracted wave converged at designed focal point (2.24$\lambda$ above the meta-device) at 2.69 GHz. **d** The simulated $|E_y|^2$ distributions on the *xy* plane with $z = 2.24\lambda$, with the dashed-line circle defining the size of the focal spot. All field values are normalized against the maximum value in the corresponding pattern.

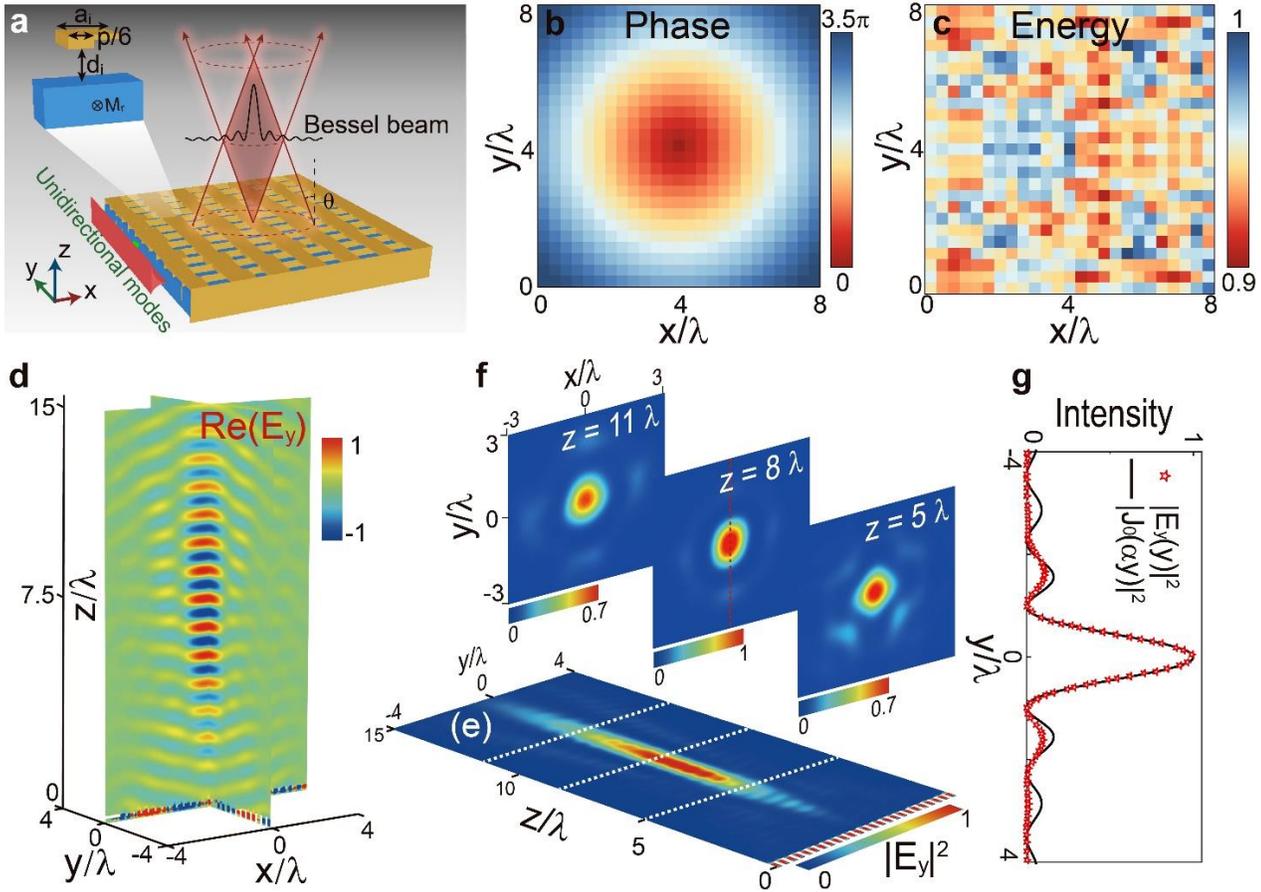

**Fig. 4| Demonstration of free-space Bessel-beam generation with a unidirectional guided-wave-driven metasurface. a** Schematic diagram of the USMP-driven meta-axicon which is composed of L-shaped USMP cells. The scattering wave from unidirectional waveguide is deflected to an angle $\theta$ toward its center to generate Bessel beam. **b**, **c** The simulated phase and energy of the extracted wave from each USMP cell of the designed meta-axicon. **d** Simulated 3D Re($E_y$) distribution for our meta-device (placed at $z = 0$ mm) driven by unidirectional guided-wave at 2.69 GHz. **e** Normalized intensity profiles $|E_y|^2$ of the generated Bessel beam in $yz$ plane with $x = 0$ and **f** $xy$ planes with $z = 5\lambda$, $8\lambda$, and $11\lambda$. **g** Normalized $|E_y|^2$ distributions along the line with $z = 8\lambda$ and $x = 0$, and the comparison with theoretical formula for a zero-order Bessel functions $J_0(\alpha y)$.

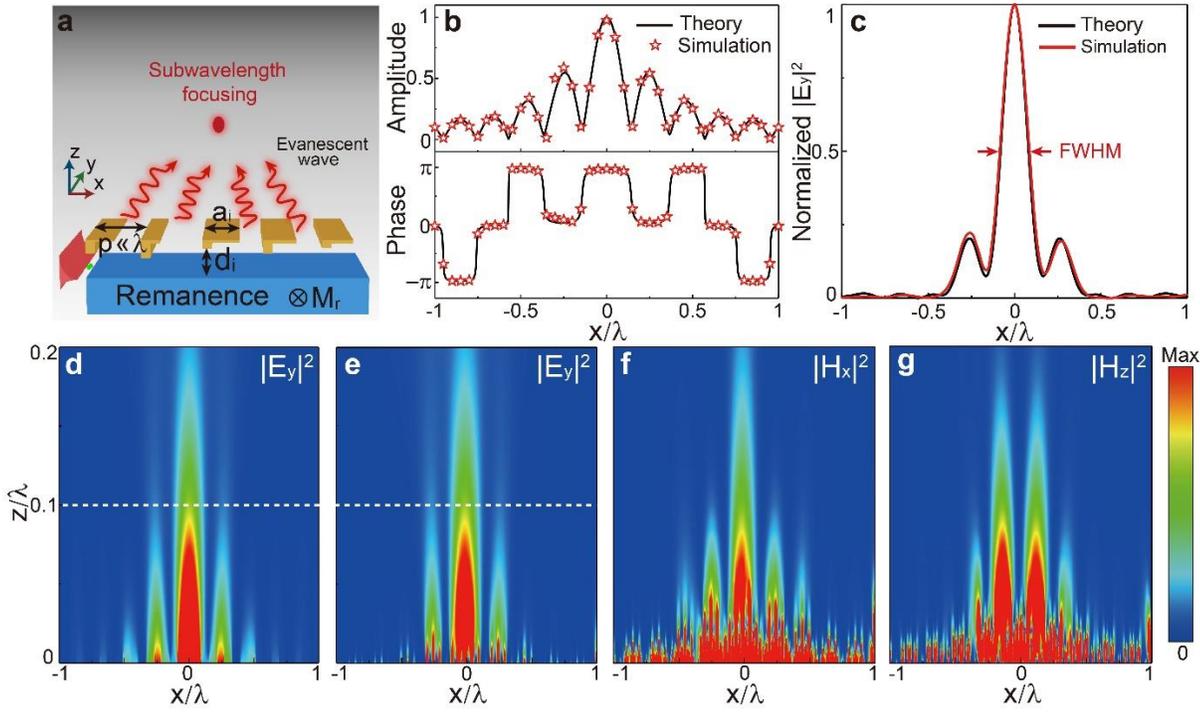

**Fig. 5| Demonstration of *sub-diffraction* focusing with a unidirectional guided wave-driven metasurface. a** Schematic diagram of the USMP-driven perfect lens which is composed of 2D L-shaped USMP cells with a deep subwavelength size $p \ll \lambda$. The extracted wave contains not only propagating wave but also evanescent wave, and the sub-diffraction focusing can be realized with the contribution of evanescent modes. **b** Designed amplitude, phase profiles localized in $[-\pi, \pi]$, and the simulated amplitude and phase of the extracted wave from each USMP cell. **c** Normalized $|E_y|^2$ at the focal line $z = 0.1\lambda$, showing a hotspot size of less than $\lambda/5$, along with the theoretical $|E_y|^2$ profile. **d** Theoretically computed field pattern $|E_y|^2$ excited by an incident wavefront $E_y = U_0(x)$ of finite size $2\lambda$ at $z = 0$. **e-g** Normalized intensity profiles $|E_y|^2$, $|H_x|^2$ and $|H_z|^2$ of the generated sub-diffraction focusing beam.

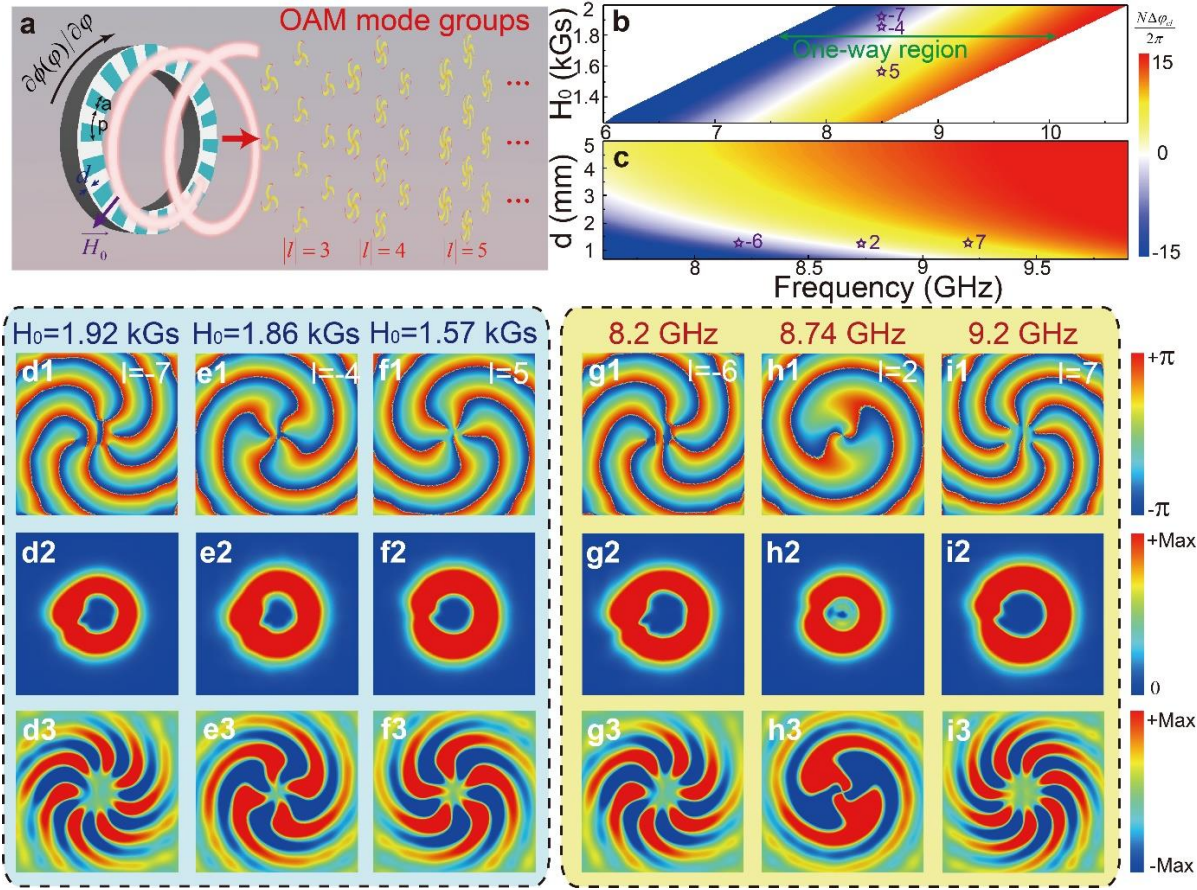

**Fig. 6| Tunable ring-cavity OAM source based on the unidirectional guided wave-driven metasurface. a** Schematic of a ring-cavity OAM source enabled by the unidirectional guided wave-driven metasurface. Unidirectional phase modulation provided by the metasurface intrinsically breaks the degeneracy of CW and CCW WGMs, leading to multiple OAM radiation at different frequencies. The dependence of USMP dispersion with geometric parameter and external magnetic field makes the radiated OAM modes tunable. **b**, **c** Numerical characterization of $N\Delta\varphi_{cl}/2\pi$ from the ring-cavity OAM source with varied external magnetic field ($H_0$) and dielectric depth ($d$) at different frequencies, the integer value of which corresponds to OAM with certain topological charge. **d1**-**f3** Simulated far-field phase, intensity and real part of $E_r$ profiles of generated beams tuned through external magnetic field ($H_0$) corresponding to different OAM eigenstates, which are marked in (**b**). **g1**-**i3** The corresponding results of generated OAM beams with different eigenstates for different frequencies, which are marked in (**c**).